\documentclass[twocolumn,secnumarabic,amssymb, nobibnotes, aps, prd]{revtex4-2}

\usepackage[colorlinks=true,linkcolor=blue]{hyperref}%
\expandafter\ifx\csname package@font\endcsname\relax\else
 \expandafter\expandafter
 \expandafter\usepackage
 \expandafter\expandafter
 \expandafter{\csname package@font\endcsname}%
\fi
\usepackage{graphics}
\usepackage{graphicx}
\usepackage{dcolumn}
\usepackage{epsfig}
\usepackage{subfigure}

\usepackage{verbatim}
\usepackage{amsmath}
\usepackage{amssymb}
\usepackage{color}
\usepackage{bm}  
\usepackage{hyperref}
 \usepackage{stackrel}
 \usepackage{upgreek}
 \usepackage{csquotes}

\def\##1{{\bf #1}}
\def\=#1{\underline{\underline #1}}
\def\^#1{{\rR}eve{#1}}
\def\`#1{{#1^\prime}}
\def\:#1{#1^{\prime\prime}}

\def\le{\left(}
\def\ri{\right)}
\def\les{\left[}
\def\ris{\right]}
\def\lec{\left\{}
\def\ric{\right\}}

 \def\eps{\varepsilon}

 \def\epso{\eps_{\scriptscriptstyle 0}}
\def\lambdao{\lambda_{\scriptscriptstyle 0}}
\def\muo{\mu_{\scriptscriptstyle 0}}
\def\ko{k_{\scriptscriptstyle 0}}
\def\etao{\eta_{\scriptscriptstyle 0}}

\def\epsa{\eps_{\rm a}}
\def\epsb{\eps_{\rm b}}
\def\epsc{\eps_{\rm c}}

 \def\.{\mbox{ \tiny{$^\bullet$} }}

\def\ux{\hat{\#x}}
\def\uy{\hat{\#y}}
\def\uz{\hat{\#z}}

\def\aL{a_{\rm L}}
\def\aR{a_{\rm R}}
\def\rL{r_{\rm L}}
\def\rR{r_{\rm R}}
\def\tL{t_{\rm L}}
\def\tR{t_{\rm R}}

\def\tLL{t_{\rm LL}}
\def\tLR{t_{\rm LR}}
\def\tRL{t_{\rm RL}}
\def\tRR{t_{\rm RR}}

 \def\TL{T_{\rm L}}
\def\TR{T_{\rm R}}

\def\thetainc{\theta_{\rm inc}}

\def\alphainc{\alpha^{\rm inc}}
\def\betainc{\beta^{\rm inc}}
\def\bphit{\les{\underline \phi}\ris}
\def\bphitinc{\les{\underline \phi}^{\rm inc}\ris}

\def\bphitincL{\les{\underline \phi}^{\rm inc}_{\rm L}\ris}
\def\bphitincR{\les{\underline \phi}^{\rm inc}_{\rm R}\ris}
\def\strazero{s_0^{\rm tr}}
\def\straone{s_1^{\rm tr}}
\def\stratwo{s_2^{\rm tr}}
\def\strathree{s_3^{\rm tr}}
\def\alphatra{\alpha^{\rm tr}}
\def\betatra{\beta^{\rm tr}}
\def\bphittra{\les{\underline \phi}^{\rm tr}\ris}

\def\bphittraL{\les{\underline \phi}^{\rm tr}_{\rm L}\ris}
\def\bphittraR{\les{\underline \phi}^{\rm tr}_{\rm R}\ris}

\def\bPhi{\Phi}

\def\bPhiL{ \Phi_{\rm L}}
\def\bPhiR{\Phi_{\rm R}}

\def\Arg{{\rm Arg}}

\def\LCSTF{L_{\rm CSTF}}
\def\LCTF{L_{\rm CTF}}
\def\twist{\varphi_{\rm t}}

\def\sp{\#s}
\def\pinc{\#p_+}
\def\pref{\#p_-}

\def\Ei{\#E^{\rm inc}(\#r)}
\def\Hi{\#H^{\rm inc}(\#r)}
\def\Er{\#E^{\rm ref}(\#r)}
\def\Hr{\#H^{\rm ref}(\#r)}
\def\Et{\#E^{\rm tr}(\#r)}
\def\Ht{\#H^{\rm tr}(\#r)}

\begin{document}

\title{Geometric-phase signature of a  structurally chiral dielectric slab with a central phase defect}%

\author{Akhlesh Lakhtakia}%
\email{akhlesh@psu.edu}
\affiliation{Department of Engineering Science and Mechanics, The Pennsylvania State University,
1450 White Course Drive, University Park, PA 16802, USA}%

\date{\today}%

\begin{abstract}
A slab made of a dielectric structurally chiral medium (DSCM) strongly reflects the co-handed
circularly polarized plane wave, but not the cross-handed circularly polarized plane wave, in
a spectral regime called
the circular Bragg regime. 
The effect of inserting a central phase defect in a DSCM slab with a modest number
of structural periods is 
a spectral reflection hole 
in the circular Bragg regime, for co-handed incidence only.
However, if the incident plane wave is left-circularly polarized, the geometric phase
of the  transmitted plane wave contains evidence of both the circular Bragg regime and the
spectral reflection hole, 
regardless
of the structural handedness of the DSCM. This evidence is indicative of the type of phase defect. 
The effect of inserting a central phase defect in a DSCM slab with a large number
of structural periods is 
a spectral transmission hole 
in the circular Bragg regime, for cross-handed incidence only.
The  spectral transmission hole may be difficult to observe experimentally because of absorption inside the DSCM slab, but it will still be
evident in  the geometric phase
of the  transmitted plane wave, if the incident plane wave is left-circularly polarized.
\end{abstract}

\maketitle

\section{Introduction}

A rectilinearly propagating plane wave is partially reflected by and partially transmitted through a slab made of 
a homogeneous isotropic dielectric material. When that material is periodically non-homogeneous
in the thickness direction, the slab  exhibits  spectral regimes of high
reflectance and correspondingly low transmittance, depending on the direction of propagation of the incident plane wave,
provided the slab thickness is several periods \cite{Bragg,Ewald}. 
The slab is often called a scalar Bragg grating and the high--reflectance spectral regimes are called
the Bragg regimes (or zones). The focus of this paper is on the lowest-order Bragg regime.

When a central phase defect is inserted in a scalar Bragg grating, the (lowest-order) Bragg  regime is punctured by
a  much narrower high--transmittance spectral regime, as demonstrated by Haus and Shank in 1976 \c{Haus}.
This narrower regime is
called a  spectral reflection hole  and is widely employed in
laser optics \cite{Agra1} and optical--fiber communication
\cite{Agra2, Bak}. Commonly, the central phase defect is a thin slab of a homogeneous dielectric material \cite{Marz,Iizuka,Baumeister}.

 In order to realize a
circular--polarization--sensitive spectral reflection hole,  the slab must be made of a dielectric structurally chiral material (DSCM),
exemplified by chiral sculptured thin films and and chiral liquid crystals \cite{LM-layer,HWLM-layer,HWTLM-twist,LVM,Vetrov2014,Pyatnov2018},
and must be thick enough to have a moderate number of structural periods.
In general, a DSCM slab
discriminates between incident plane waves of different circular
polarization states in the Bragg regime. DSCMs and circularly polarized plane waves
possess handedness.
In the Bragg regime,
the reflectance of a DSCM slab is very high
for a co--handed  incident plane wave, but not for the
cross--handed one~---~leading to the term {\em circular Bragg regime} \cite{FLaop}.
As the high reflectance in the circular Bragg regime is only for the
co--handed incident plane wave, so is the spectral reflection hole arising from 
 the insertion of a central phase defect in the DSCM slab.

Facile implementation of central phase defects in a DSCM slab is
possible of the following two types:
\begin{itemize}
\item[(i)] Layer defect: A homogeneous layer, whether isotropic \cite{HWLM-layer} or anisotropic \cite{LVM}, of finite thickness
is inserted in the center of the DSCM slab. The thickness of the homogeneous layer determines the
center wavelength of the spectral reflection hole.
\item[(ii)] Twist defect: One half of the DSCM slab is twisted about the thickness
axis with respect to the other half by a certain angle \cite{HWTLM-twist,Pyatnov2018}. The   twist angle
determines the center--wavelength of the spectral reflection hole.
\end{itemize}
Both types of phase defects may be combined to offer
 design flexibility \cite{Vetrov2014}. Needless to add, a twist defect is not possible with scalar Bragg gratings.
 
DSCMs without a phase defect are attractive as  relatively wideband optical filters  \cite{Adams,Jacobs,Park2008,Huang2014,Chen2015,Fiallo},
 whereas DSCMs with a central phase defect are attractive as narrowband optical filters \cite{HWTLM-twist,Pursel2007,Pursel-OE}.
 Both types of filters are also  deployed for optical sensing \cite{Liu2011,Pursel2007,LMSWH,Lee2016,Chen2021} and lasing \cite{Schmidtke-PRL,Palffy,Ford,Topf}.
 Notably, these applications are based on high or low values of reflectances and transmittances, which are all positive real numbers,
 but not on the reflection and transmission coefficients, which are complex numbers \cite{STFbook}.
 
 A quantity derivable from the transmission coefficients is the geometric phase \cite{Panchu,Bhandari}. This quantity is  of current interest 
 for research on chiral liquid crystals \cite{Barboza,Brass2016,Kobashi,Chen2020} and chiral sculptured thin films \cite{Das,Lakh2024},
 with some potential for application \cite{Brass2016,Chen2020,Jisha}. The geometric-phase spectrum of    plane-wave
 transmission through a DSCM slab contains a signature of the circular Bragg phenomenon,
provided that the incident plane wave is not right circularly polarized, whether the DSCM is structurally right handed or structurally left handed \cite{Lakh2024}
Furthermore,  both the thickening of the DSCM slab and the reversal of structural handedness affect that geometric phase.

This paper is devoted to the signatures of central phase defects in the geometric-phase spectrum of the transmitted plane wave.
The  phase defect can be either (i) an anisotropic layer defect, (ii) a twist defect, or (iii) a combination of both defects. For calculations,
the DSCM
is taken to be a chiral sculptured thin film \cite{Nature1,Nature2} and the anisotropic layer defect to be a columnar thin film \cite{HWbook}. Without significant loss
of generality, both   films are taken to be sequentially made by evaporating the same material in a vacuum chamber and the substrate is
 similarly oriented for the growth of both films, the only difference being that the substrate spins about a central normal
axis passing through it when a chiral sculptured thin film is being grown but is stationary when a columnar thin film is being grown \cite{STFbook,HWbook}.

This paper is organized as follows. Section~\ref{sec:theory} provides 
the theoretical framework to calculate  the geometric phase of the transmitted plane wave
in relation to the incident plane wave. Numerical results are presented
and discussed in
Sec.~\ref{sec:nrd}, and the paper ends with key conclusions in Sec.~\ref{sec:cr}.
An $ \exp(-i \omega t)$  dependence on time $t$ is explicit, with $ \omega=2\pi{f}$ as the angular frequency,
$f$ as the linear frequency,
and $ i = \sqrt{-1}$.  With $ \epso$ and $\muo$, respectively,
denoting the permittivity and permeability  of free space,
the free-space wavenumber is denoted by $\ko 
= \omega \sqrt{\epso \muo}$, $\lambdao=2\pi/\ko$ is the free-space
wavelength, and $\etao=\sqrt{\muo/\epso}$ is the
intrinsic impedance of free space. The Cartesian coordinate system $(x,y,z)$ is adopted.
Vectors are in boldface
 and unit vectors are additionally decorated by a caret on top. Dyadics are double underlined.
Column vectors are underlined and enclosed in square brackets. 
 The asterisk $^\ast$ denotes the
complex conjugate, and the dagger $^\dag$ the conjugate transpose.

\section{Theory} \label{sec:theory}

\begin{figure}[!ht]
\begin{center}
\begin{tabular}{c}
\includegraphics[width=0.25\linewidth]{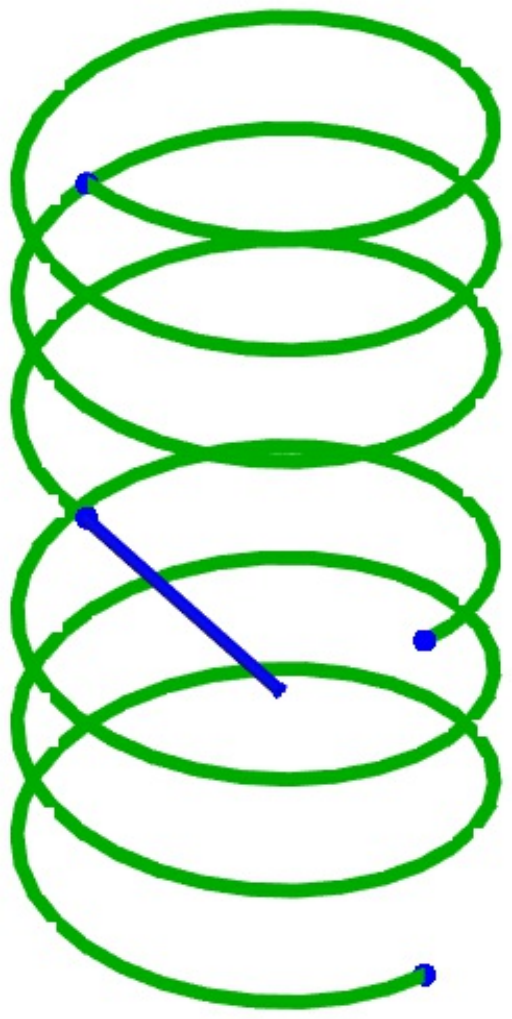}
\includegraphics[width=0.25\linewidth]{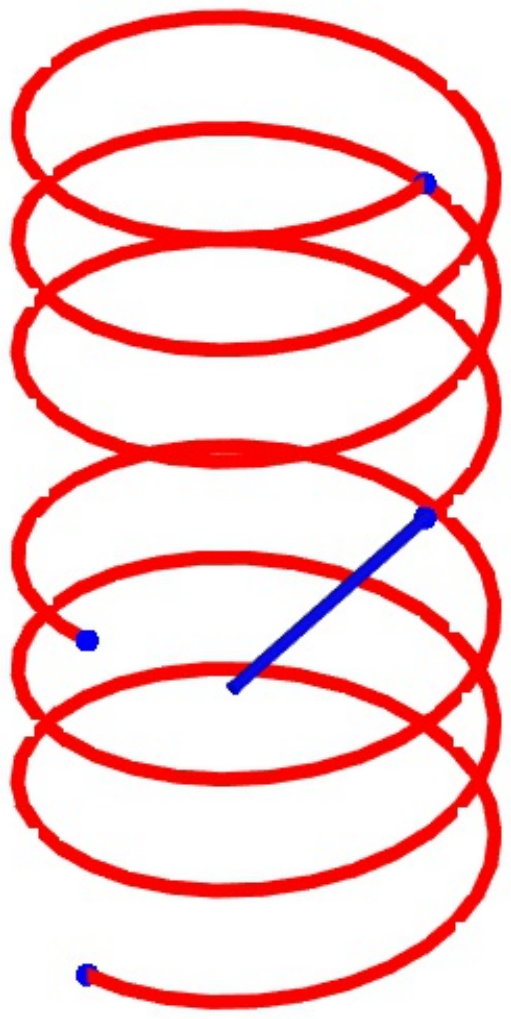}
\end{tabular}
\end{center}
\caption{\label{Chiral}
Morphological cartoons for (left)  left-handed  and (right)
right-handed chiral sculptured thin films with a central layer defect
and a central twist defect.
}
\end{figure}

The DSCM slab with a central layer defect and a central twist defect occupies the region $0<z<L$,
where the regions $0<z<\LCSTF=2N\Omega$
and $\LCSTF+\LCTF<z<L=2\LCSTF +\LCTF$ are occupied by a chiral sculptured thin film and
the region $\LCSTF<z<\LCSTF+\LCTF$ by a columnar thin film,
captured by the schematics shown in Fig.~\ref{Chiral}.
 Here, $2\Omega$ is the period
of the chiral sculptured thin film along the $z$ axis and $N\in\lec1,2,3,\dots\ric$ is the number
of periods on either side of the central phase defect.

\subsection{Relative Permittivity Dyadic}
The relative permittivity dyadic of the dielectric matter in the region $0<z<L$
is given by
\begin{eqnarray}
 \nonumber
&&\=\eps_{\rm rel}(z)=
\=S_{\rm z}(h,p,\varphi,z) \.
\=S_{\rm y}(\chi)\.\left[\epsa\uz\uz+\epsb\ux\ux
+\epsc\uy\uy\right]
\nonumber
\\[5pt]
&&\qquad 
\.\=S_{\rm y}^{-1}(\chi) \.\=S_{\rm z}^{-1}(h,p,\varphi,z)\,,
\quad z\in(0,L)\,.
\label{epsDSCM}
\end{eqnarray}
 The frequency-dependent relative permittivity scalars
$\epsa$, $\epsb$, and $\epsc$ capture   local orthorhombicity \cite{Kittel,STFbook,HWbook}.
 The tilt dyadic
\begin{eqnarray}
\nonumber
 &&
\=S_{\rm y} (\chi) = \uy\uy + \left(\ux\ux+\uz\uz\right)\cos\chi
 \\
 &&\qquad
 + \left(\uz\ux-\ux\uz\right)\sin\chi
 \label{SY-def}
 \end{eqnarray}
contains $\chi\in[0\ \text{deg},90\ \text{deg}]$ as an angle of inclination with respect to the $xy$ plane.

Both structural handedness and twist are captured
by the rotation dyadic
\begin{eqnarray}
\nonumber
&&
\=S_z(h,p,\varphi,z)=\uz\uz +\left(\ux\ux+\uy\uy\right)\cos\les{h\left({\pi}pz+\varphi\right)}\ris
\\
&&
\qquad
+\left(\uy\ux-\ux\uy\right)\sin\les{h\left({\pi}pz+\varphi\right)}\ris\,.
\label{Sz-def}
\end{eqnarray} 
Here,
$h \in\left\{-1,1\right\}$ is the structural-handedness parameter, with $h=-1$ for structural left-handedness and
 $h=1$ for structural right-handedness; $p/2$ is the reciprocal of the period;
 and $\varphi\in[0,2\pi)$ is a twist about the $z$ axis.
 The latter two parameters are defined in piecewise fashion as follows:
 \begin{eqnarray}
 \nonumber
 &&p=\left\{\begin{array}{c} 1/\Omega \\ 0 \\ 1/\Omega\end{array}\right.\,,\quad
 \varphi =\left\{\begin{array}{c} 0 \\ 0 \\ \twist \end{array}\right.\,,
 \\[8pt]
 &&\qquad
z\in \left\{ \begin{array}{c} (0,\LCSTF) \\ (\LCSTF,\LCSTF+\LCTF) \\ (\LCSTF+\LCTF,L) \end{array}\right.\,.
\end{eqnarray}

The foregoing equations also apply for a chiral smectic liquid crystal containing a layer of a smectic liquid crystal \cite{deG,Parodi}. 
Furthermore, the same equations
can  be used for a chiral nematic liquid crystal containing a layer of a nematic
liquid crystal by setting $\epsc=\epsa$ and $\chi=0$ \cite{deG}. It is also possible to specify $\varepsilon_{\rm a,b,c}$ and $\chi$
in the  defect layer   differently from their respective values in the regions $0<z<\LCSTF=2N\Omega$
and $\LCSTF+\LCTF<z<L=2\LCSTF +\LCTF$, but that  is not necessary for the present purpose.

\subsection{Boundary-Value Problem}
The half-space $z < 0$ is the region of incidence and reflection, while
the half-space $z > L$ is the region of transmission.
A plane
wave, propagating in the half--space
$z \leq 0$ at an angle $\thetainc\in[0,\pi/2)$ to the $z$ axis and at an angle $\psi\in[0,2\pi)$
to the $x$ axis in the $xy$ plane, is incident on the chosen chiral STF.
The electric and magnetic field phasors associated
with the incident plane wave are represented as \cite{STFbook}
\begin{subequations}
\begin{eqnarray}
\nonumber
&&\Ei= \les \frac{\le i\sp - \pinc \ri}{\sqrt{2}} \, \aL -\, \frac{\le i\sp +\pinc
\ri}{\sqrt{2}} \, \aR \ris \\
\nonumber
& & \times \, \exp{\les i\kappa \le x\cos\psi + y\sin\psi \ri
\ris} \exp{\left(i\ko z{\cos\thetainc}\right)} \, , 
\\
\label{incEphas}
&&\qquad   z < 0 \, ,
\end{eqnarray}
and
\begin{eqnarray}
\nonumber
&&\Hi= \frac{1}{i\etao}\les \frac{\le i\sp - \pinc \ri}{\sqrt{2}} \, \aL +\,
\frac{\le i\sp + \pinc \ri }{\sqrt{2}} \, \aR \ris \\
\nonumber
& & \times \, \exp{\les i\kappa \le x\cos\psi + y\sin\psi \ri
\ris} \exp{\left(i\ko z{\cos\thetainc}\right)} \, ,
\\
&& \qquad z
< 0\,,
\label{incHphas}
\end{eqnarray}
\end{subequations}
where
 \begin{equation}
 \left.\begin{array}{l}
\kappa =
\ko\sin{\thetainc}\\[5pt]
\sp=-\ux\sin\psi + \uy \cos\psi \\[5pt]
\#p_\pm=\mp\le \ux \cos\psi + \uy \sin\psi \ri \cos{\thetainc} 
 + \uz \sin{\thetainc}
\end{array}
\right\}
\, .
\end{equation}
The amplitudes of the left-circularly polarized (LCP) and the 
right-circularly polarized (RCP) components of the incident plane wave, denoted by 
$\aL$ and
$\aR$, respectively, are assumed to be known.

The reflected electric and magnetic field phasors are expressed as
\begin{subequations}
\begin{eqnarray}
\nonumber
&&\Er=-\, \les \frac{\le i\sp - \pref \ri}{\sqrt{2}} \, {\rL} -\, \frac{\le i\sp +
\pref \ri}{\sqrt{2}} \, {\rR} \ris \\
\nonumber
& &  \times \, \exp{\les i\kappa \le x\cos{\psi} + y\sin{\psi} \ri
\ris} \exp\left({-i\ko   z\cos{\thetainc}}\right) \, ,
\\
&& \qquad z < 0 \, ,
\label{refEphas}
\end{eqnarray}
and
\begin{eqnarray}
\nonumber
&&\Hr=-\, \frac{1}{i\etao}\les \frac{\le i\sp - \pref \ri}{\sqrt{2}} \, {\rL} +\,
\frac{\le i\sp + \pref \ri}{\sqrt{2}} \, {\rR} \ris \\
\nonumber
& & \times \, \exp{\les i\kappa \le x\cos{\psi} + y\sin{\psi} \ri
\ris} \exp\left({-i\ko   z\cos{\thetainc}}\right) \, ,  
\\ 
&&\qquad z < 0 \, .
\label{refHphas}
\end{eqnarray}
\end{subequations}
The transmitted electric and magnetic field phasors are represented as
\begin{subequations}
\begin{eqnarray}
\nonumber
&&
\Et= \les \frac{\le i\sp - \pinc \ri}{\sqrt{2}} \, {\tL} -\, \frac{\le i\sp +\pinc
\ri}{\sqrt{2}} \, {\tR} \ris \\
\nonumber
& &  \times \, \exp{\les i\kappa \le x\cos{\psi} + y\sin{\psi} \ri
\ris} \exp\les{i\ko   (z-L)\cos{\thetainc}}\ris \, , 
\\&&\qquad z > L \, ,
\label{trEphas}
\end{eqnarray}
and
\begin{eqnarray}
\nonumber
&&
\Ht= \frac{1  }{i\etao}\les \frac{\le i\sp - \pinc \ri}{\sqrt{2}} \, {\tL} +\,
\frac{\le i\sp + \pinc \ri }{\sqrt{2}} \, {\tR} \ris \\
\nonumber
& & \times \, \exp^{\les i\kappa \le x\cos{\psi} + y\sin{\psi} \ri
\ris} \exp\les{i\ko   (z-L)\cos{\thetainc}}\ris \, , 
\\&&
\qquad z
> L
\, .
\label{trHphas}
\end{eqnarray}
\end{subequations}

The reflection amplitudes (${\rL}$ and ${\rR}$) as well as the transmission
amplitudes (${\tL}$ and ${\tR}$) are unknown and require the solution
of a boundary-value problem. Several numerical techniques exist to solve
this problem \cite{Dreher,Sugita, Oldano,LW95}. The most straightforward
technique requires the use of the piecewise uniform approximation of 
$\=\eps_{\rm rel}(z)$ followed by application of the 4$\times$4 transfer-matrix method
\cite{MLbook}. The interested reader is
referred to  Ref.~\citenum{STFbook} for a detailed description of this technique.

Thereafter, the transmittance
\begin{equation}
T=\frac{\vert\tL\vert^2+\vert\tR\vert^2}{\vert\aL\vert^2+\vert\aR\vert^2}\,\in[0,1]
\end{equation}
can be calculated.

\subsection{Geometric Phase}

Any plane wave can be located on the Poincar\'e sphere using the
polar angle $\alpha\in[0,2\pi)$ and the azimuthal angle $\beta\in[-\pi/2,\pi/2]$,
and its Poincar\'e spinor can then be defined as
\begin{equation}
\bphit = \les\begin{array}{c}
\cos\left(\frac{\pi}{4}-\frac{\beta}{2}\right)
\\[5pt]
\sin\left(\frac{\pi}{4}-\frac{\beta}{2}\right) \exp\left(i\alpha\right)
\end{array}
\ris\,.
\end{equation}
The transmitted plane wave possesses the geometric  phase
\begin{equation}
\bPhi=\Arg\lec\bphitinc^\dag\.\bphittra\ric\,
\end{equation}
relative to the incident plane wave.

For an incident LCP plane wave, $\alphainc=0$ and $\betainc=-\pi/2$ so
that 
\begin{equation}
\bphitincL = \les\begin{array}{c}
0
\\[5pt]
1
\end{array}
\ris\,
\end{equation}
is the Poincar\'e spinor. 
For an incident RCP plane wave, $\alphainc=0$, $\betainc=\pi/2$,
and
\begin{equation}
\bphitincR = \les\begin{array}{c}
1
\\[5pt]
0
\end{array}
\ris\,
\end{equation}
is the Poincar\'e spinor.

The Stokes parameters of the transmitted plane wave are given by \cite{Jackson}
\begin{equation}
\label{Stokes-tra}
\left.\begin{array}{l}
\strazero = \vert\tR\vert^2+\vert\tL\vert^2
\\[5pt]
\straone =2\,{\rm Re}\left(\tL\,\tR^\ast\right)
\\[5pt]
\stratwo =2\,{\rm Im}\left(\tL\,\tR^\ast\right)
\\[5pt]
\strathree = \vert\tR\vert^2-\vert\tL\vert^2
\end{array}
\right\}\,.
\end{equation} 
Accordingly, the angles $\alphatra$
and $\betatra$ can be  calculated using 
\begin{equation}
\left.\begin{array}{l}
\straone =\strazero \cos\betatra \cos\alphatra
\\[5pt]
\stratwo =\strazero \cos\betatra \sin\alphatra
\\[5pt]
\strathree=\strazero\sin\betatra
\end{array}
\right\}\,,
\end{equation}
so that 
\begin{equation}
\bphittra = \les\begin{array}{c}
\cos\left(\frac{\pi}{4}-\frac{\betatra}{2}\right)
\\[5pt]
\sin\left(\frac{\pi}{4}-\frac{\betatra}{2}\right)  \exp\left(i\alphatra\right)
\end{array}
\ris
\end{equation}
is the Poincar\'e spinor of the transmitted plane wave. 
For later convenience, let $\alphatra_{\rm L}$, $\betatra_{\rm L}$, and $\bphittraL$
refer to LCP incidence, whereas $\alphatra_{\rm R}$, $\betatra_{\rm R}$, and $\bphittraR$
refer to RCP incidence.

\section{Numerical Results and Discussion}\label{sec:nrd}

Single-resonance Lorentzian dependences on $\lambdao$ were assumed 
for $\epsa$, $\epsb$, and $\epsc$
as follows \cite{Kittel}:
\begin{equation}
\label{resonance}
\eps_{\rm a,b,c}(\lambdao) = 1+ \frac{p_{\rm a,b,c}}
{1 + (1/N_{\rm a,b,c}  - i \lambda_{\rm a,b,c}/ \lambdao )^2}\,.
\end{equation}
The oscillator strengths are determined by the values of $p_{\rm a,b,c}$,  $\lambda_{\rm a,b,c} (1 + N_{\rm a,b,c}^{-2})^{-1/2}  $  are the  resonance wavelengths, and $\lambda_{\rm a,b,c}/N_{\rm a,b,c}$ are the  resonance linewidths. 
 The parameters used for most of the theoretical results reported here are as follows:  $p_{\rm a} = 2.3$, $p_{\rm b} =3.0$, $p_{\rm c} =2.2 $, $\lambda_{\rm a} = \lambda_{\rm c} =260$~nm, $\lambda_{\rm b} = 270$~nm,   and
$N_{\rm a} = N_{\rm b} =N_{\rm c}=130$. Furthermore, $\chi = 37$~deg  and $\Omega = 150$~nm were fixed. 

Calculations of the transmittance $\TR$ were made by setting
 $\aR=1$ and $\aL=0$. Note that the corresponding geometric phase
\begin{equation}
\bPhiR=\Arg\lec\bphitincR^\dag\.\bphittraR\ric\equiv 0\,,
\end{equation}
as   shown elsewhere \cite{Das}.
The transmittance $\TL$ and the geometric phase $\bPhiL=\alphatra_{\rm L}$ were calculated by setting
 $\aR=0$ and $\aL=1$.  

\subsection{DSCM slab without central phase defect}

\begin{figure}[!ht]
\begin{center}
\begin{tabular}{c}
\includegraphics[width=\linewidth]{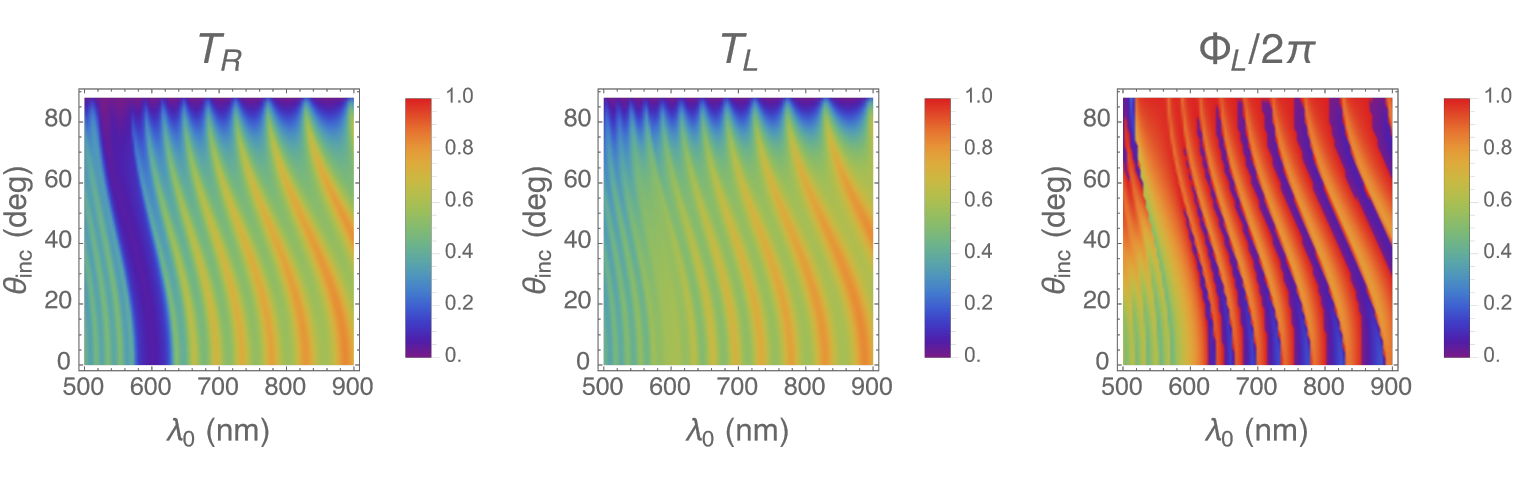}\\
\includegraphics[width=\linewidth]{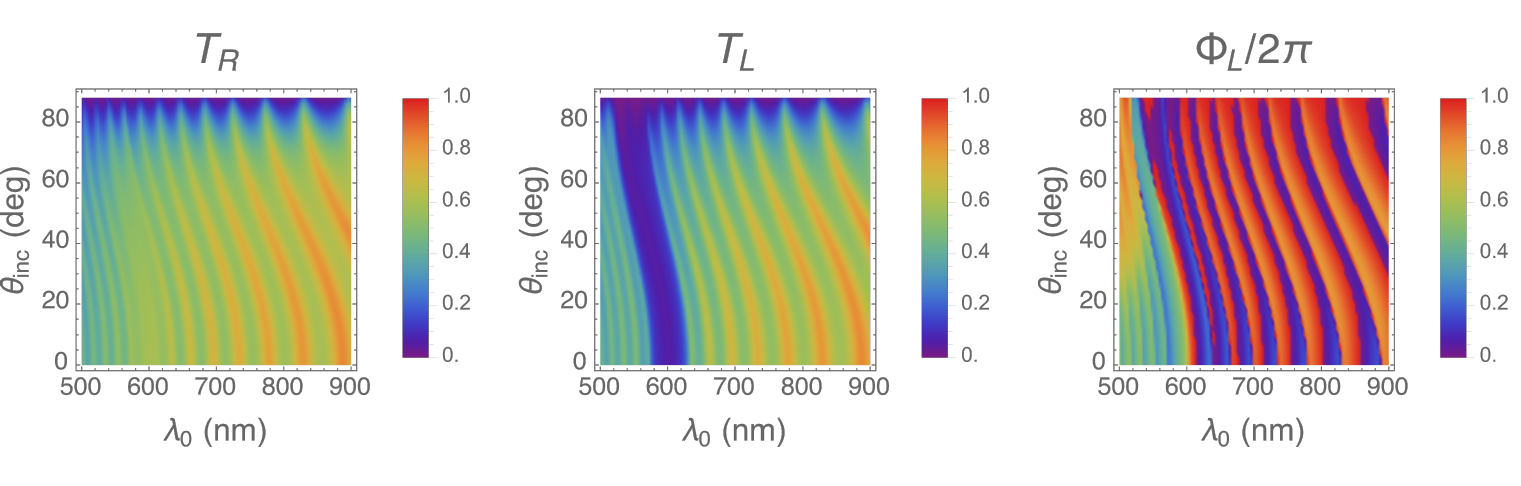}
\end{tabular}
\end{center}
\caption{\label{NoDefect}
$\TR$,  $\TL$, and $\bPhiL$ as functions
of $\lambdao$ and $\thetainc$ of a DSCM slab without a central phase
defect ($\LCTF=0$ and $\twist=0\deg$),
calculated for $\psi=0\deg$,  $\Omega=150$~nm,  and $N=5$. Other parameters are as follows: $p_{\rm a} = 2.3$, $p_{\rm b} =3.0$, $p_{\rm c} =2.2 $, $\lambda_{\rm a} = \lambda_{\rm c} =260$~nm, $\lambda_{\rm b} = 270$~nm,   and
$N_{\rm a} = N_{\rm b} =N_{\rm c}=130$.
Top row: $h = 1$. Bottom row: $h=-1$. 
}
\end{figure}

For reference, Fig.~\ref{NoDefect} presents $\TR$,  $\TL$, and $\bPhiL$ as functions
of $\lambdao\in[400~{\rm nm},900~{\rm nm}]$ and $\thetainc\in[0\deg,90\deg)$ of a DSCM slab without a
central phase defect ($\LCTF=0$ and $\twist=0\deg$),
calculated for $\psi=0\deg$, $N=5$, and $h=\pm1$ \cite{Lakh2024}. The circular Bragg regime is evident 
as a blue trough  in the plots of: (i)  $\TR$ for $h=1$ and (ii) $\TL$ for $h=-1$. The troughs are about 70-nm wide and  centered
at $\lambdao\simeq 600$~nm for $\thetainc=0\deg$;  they
 blueshift as $\thetainc$ increases,
in accord with experimental results \cite{Erten2015}. 
Although the two troughs look
identical, they are somewhat different \cite{Lakh2024}.
The blue trough is naturally absent
in the plots of $\TL$ for $h=1$ and $\TR$ for $h=-1$ \cite{FLaop}.

Whereas $\bPhiR\equiv0$, both plots of $\bPhiL$ vs. $\lambdao$ and $\thetainc$ in Fig.~\ref{NoDefect}
contain a signature of the circular Bragg phenomenon. 
A reversal of structural handedness affects but does not lead to a simple change in $\bPhiL$ \cite{Lakh2024}.

\subsection{DSCM slab with a central twist defect}

\begin{figure}[!ht]
\begin{center}
\begin{tabular}{c}
\includegraphics[width=\linewidth]{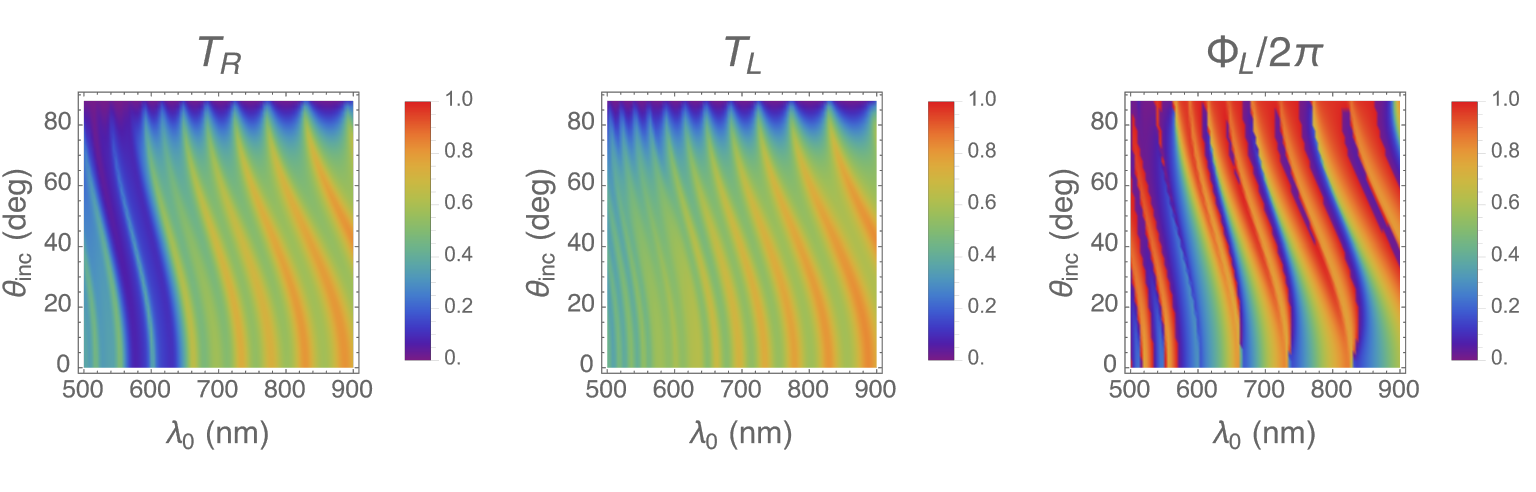}\\
\includegraphics[width=\linewidth]{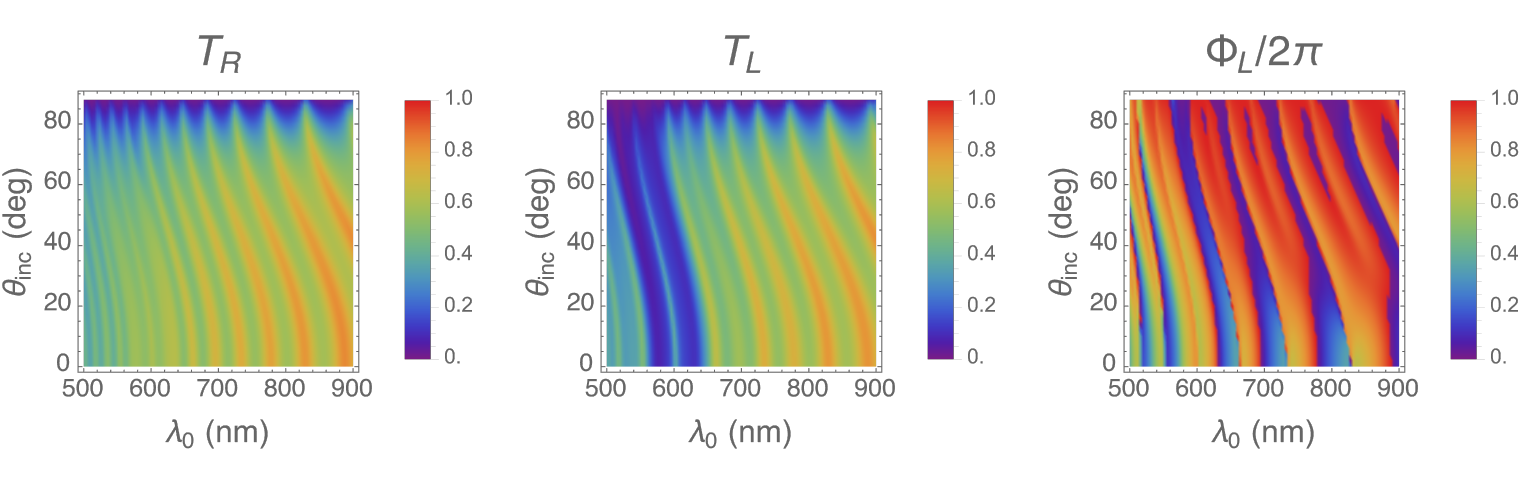}
\end{tabular}
\end{center}
\caption{\label{TwistDefect}
$\TR$,  $\TL$, and $\bPhiL$ as functions
of $\lambdao$ and $\thetainc$ of a DSCM slab with a central twist defect ($\LCTF=0$ and $\twist=90\deg$),
calculated for $\psi=0\deg$, $\Omega=150$~nm,  and $N=5$. Other parameters are as follows:
$p_{\rm a} = 2.3$, $p_{\rm b} =3.0$, $p_{\rm c} =2.2 $, $\lambda_{\rm a} = \lambda_{\rm c} =260$~nm, $\lambda_{\rm b} = 270$~nm,   and
$N_{\rm a} = N_{\rm b} =N_{\rm c}=130$.
Top row: $h = 1$. Bottom row: $h=-1$. 
}
\end{figure}

The introduction of a central $90$-$\deg$ twist defect by itself is responsible for the insertion of
a 15-nm-wide high-transmittance ridge in the center of the 70-nm-wide blue trough  in the plots of: (i)  $\TR$ vs. 
 $\lambdao\in[400~{\rm nm},900~{\rm nm}]$ and $\thetainc\in[0\deg,90\deg)$ for $h=1$ and (ii)  $\TL$
 similarly  for $h=-1$,
as illustrated in Fig.~\ref{TwistDefect} for $\psi=0\deg$.
 Like the troughs signifying the exhibition of the
 circular Bragg phenomenon, the high-transmittance ridges in both plots blueshift as $\thetainc$ increases. These ridges
 are  spectral reflection holes.

A linear feature right in the center of the
circular Bragg regime
also appears in the plots of $\bPhiL$ vs. $\lambdao$ and $\thetainc$ for $h=\pm1$ in Fig.~\ref{TwistDefect}, as a signature of the 
central twist defect.
Additionally, the red-orange-yellow curvaceous ridges on the long-wavelength side of the circular Bragg regime
in the plots of $\bPhiL$ in Fig.~\ref{NoDefect} appear to have partially coalesced  pairwise
in the plots of $\bPhiL$ in Fig.~\ref{TwistDefect}. The pairwise coalescence is partial, because it is pronounced for
low $\thetainc$ with bifurcation evident for high $\thetainc$. Again, 
a reversal of structural handedness  does not lead to a simple change in $\bPhiL$ in Fig.~\ref{TwistDefect}.

Calculations (not shown here) indicate that the   spectral reflection holes for $h=\pm1$
redshift within the
circular Bragg regime when $\twist$ is progressively reduced from $90\deg$ to  $60\deg$, and they 
 blueshift when $\twist$ is progressively increased from $90\deg$ to $120 \deg$.

\subsection{DSCM slab with a central layer defect}

\begin{figure}[!ht]
\begin{center}
\begin{tabular}{c}
\includegraphics[width=\linewidth]{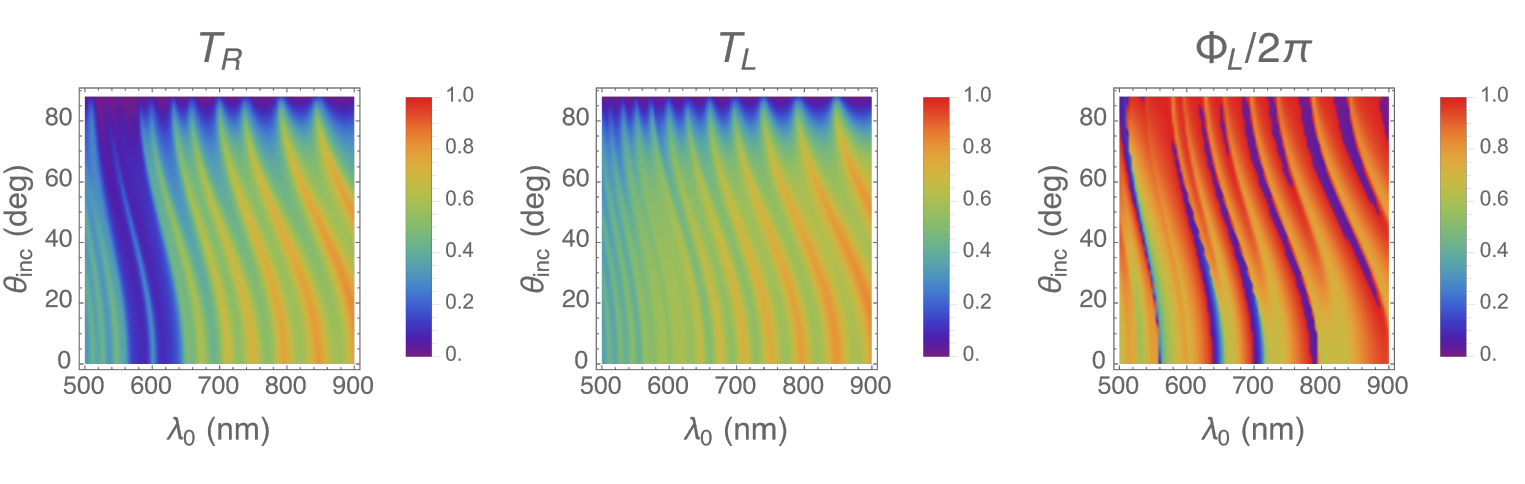}\\
\includegraphics[width=\linewidth]{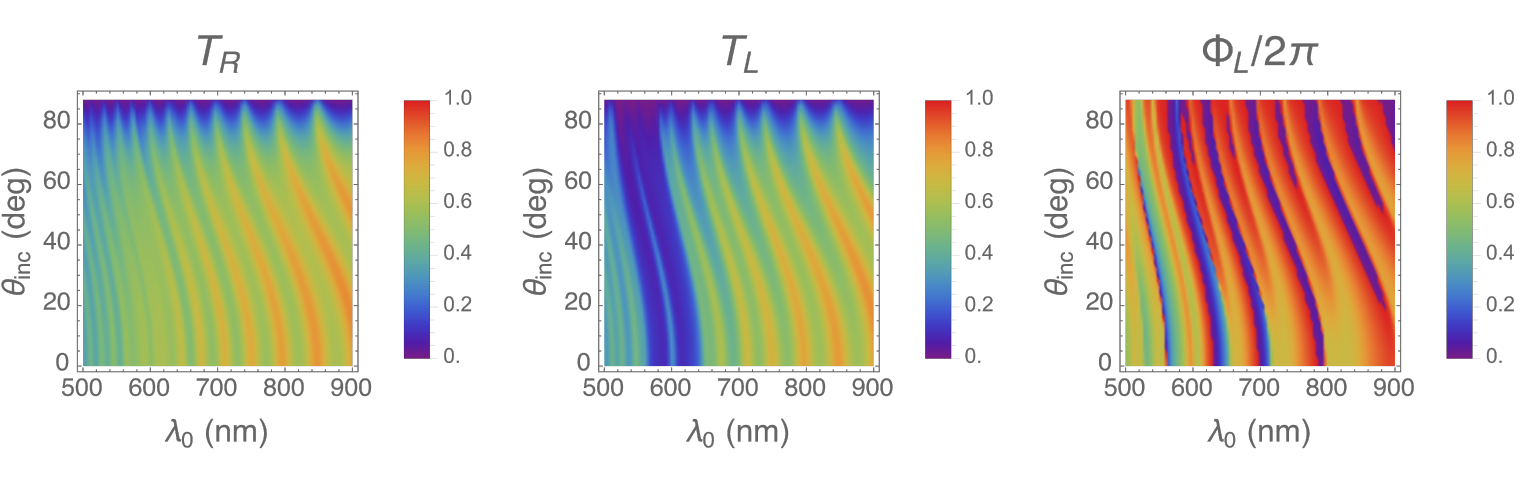}
\end{tabular}
\end{center}
\caption{\label{LayerDefect}
$\TR$,  $\TL$, and $\bPhiL$ as functions
of $\lambdao$ and $\thetainc$ of a DSCM slab with a central layer defect ($\LCTF=\Omega/2$ and $\twist=0\deg$),
calculated for $\psi=0\deg$, $\Omega=150$~nm, and $N=5$. Other parameters are as follows:
$p_{\rm a} = 2.3$, $p_{\rm b} =3.0$, $p_{\rm c} =2.2 $, $\lambda_{\rm a} = \lambda_{\rm c} =260$~nm, $\lambda_{\rm b} = 270$~nm,   and
$N_{\rm a} = N_{\rm b} =N_{\rm c}=130$.
Top row: $h = 1$. Bottom row: $h=-1$.
}
\end{figure}

A central $\Omega/4$-thick layer defect alone is responsible for the insertion of a
 central 15-nm-wide high-transmittance ridge (i.e., spectral reflection hole) in the blue  70-nm-wide trough  in the plots of: (i)  $\TR$ 
 for $h=1$ and  (ii) $\TL$ for $h=-1$,
 in Fig.~\ref{LayerDefect}.  The high-transmittance ridges blueshift as $\thetainc$ increases, just as
 in Fig.~\ref{TwistDefect}.

A linear feature  in the center of the circular Bragg regime  and partial pairwise coalescence
of curvaceous ridges on the long-wavelength side of the circular Bragg regime are also evident
 in the plots of $\bPhiL$ vs. $\lambdao$ and $\thetainc$ for $h=\pm1$ and $\psi=0\deg$ in Fig.~\ref{LayerDefect}. However, visual comparison of Figs.~\ref{TwistDefect} and \ref{LayerDefect}
reveals  
 that  the effects of the central twist defect and the central layer defect on $\bPhiL$ are quantitatively not  identical.

The spectral reflection holes blueshift  within the circular Bragg regime
when $\LCTF$ is progressively reduced from $\Omega/2$  to  $\Omega/3$,
and redshift when $\LCTF$ is progressively increased from $\Omega/2$ to $2\Omega/3$, according to calculations not reported
here in detail.

\subsection{DSCM slab with  central layer and twist defects}

\begin{figure}[!ht]
\begin{center}
\begin{tabular}{c}
\includegraphics[width=\linewidth]{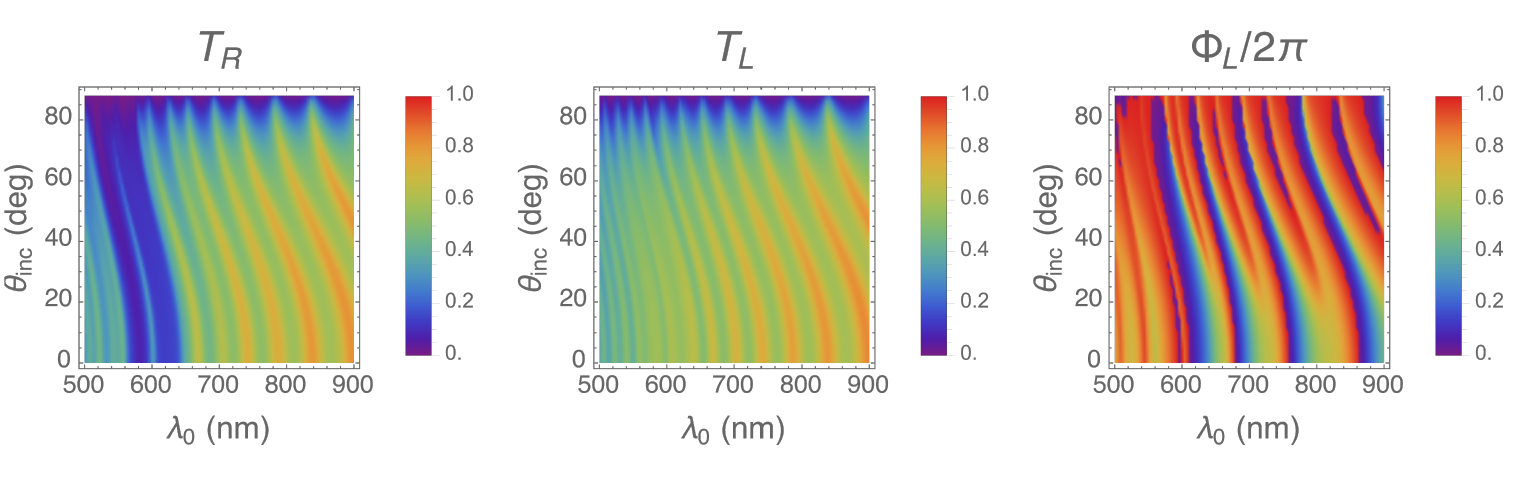}\\
\includegraphics[width=\linewidth]{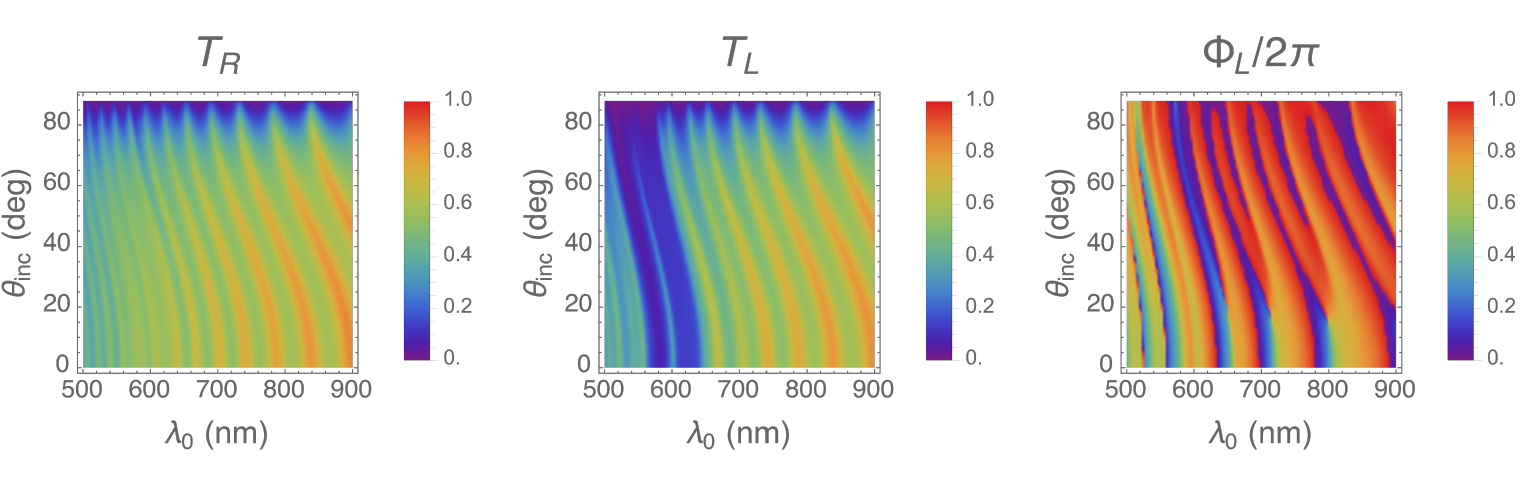}
\end{tabular}
\end{center}
\caption{\label{CombinedDefect}
$\TR$,  $\TL$, and $\bPhiL$ as functions
of $\lambdao$ and $\thetainc$ of a DSCM slab with  central layer and twist defects ($\LCTF=\Omega/4$ and $\twist=135\deg$),
calculated for $\psi=0\deg$,  $\Omega=150$~nm,  and $N=5$. Other parameters are as follows:
$p_{\rm a} = 2.3$, $p_{\rm b} =3.0$, $p_{\rm c} =2.2 $, $\lambda_{\rm a} = \lambda_{\rm c} =260$~nm, $\lambda_{\rm b} = 270$~nm,   and
$N_{\rm a} = N_{\rm b} =N_{\rm c}=130$.
Top row: $h = 1$. Bottom row: $h=-1$.
}
\end{figure}

Both types of central phase defects can cooperate to generate spectral reflection holes \cite{Vetrov2014}.
Therefore, for $\LCTF=\Omega/4$ and $\twist=135\deg$, a
central 15-nm-wide high-transmittance ridge is present in the center of the blue 7-nm-wide trough  in the plots 
in Fig.~\ref{CombinedDefect} of: (i)  $\TR$ vs. $\lambdao$ and $\thetainc$ for $h=1$ and (ii) $\TL$,
also vs. $\lambdao$ and $\thetainc$, for $h=-1$.
 More importantly in the present context, a linear feature manifests as a signature of the combined
 defects
 in the plots of $\bPhiL$ vs. $\lambdao$ and $\thetainc$ for $h=\pm1$, in that figure. The two defects also combine to effect the partial pairwise coalescence
of curvaceous ridges on the long-wavelength side of the circular Bragg regime.

Figures \ref{NoDefect}--\ref{CombinedDefect} show the spectrums of
$\TR$,  $\TL$, and $\bPhiL$ for $\thetainc\in\les0\deg,90\deg\ris$ with
$\psi=0\deg$ fixed. Both $\TR$ and $\TL$ for DSCM slabs without central phase defects
have long been known to vary weakly
with $\psi$ \cite{STFbook}. In contrast, it was recently shown \cite{Lakh2024} that $\bPhiL$ does depend
significantly on $\psi$ for a defect-free DSCM slab (i.e., $\LCTF=0$ and $\twist=0\deg$), as
confirmed by Fig.~\ref{NoDefect}.

\begin{figure}[!ht]
\begin{center}
\begin{tabular}{c}
\includegraphics[width=\linewidth]{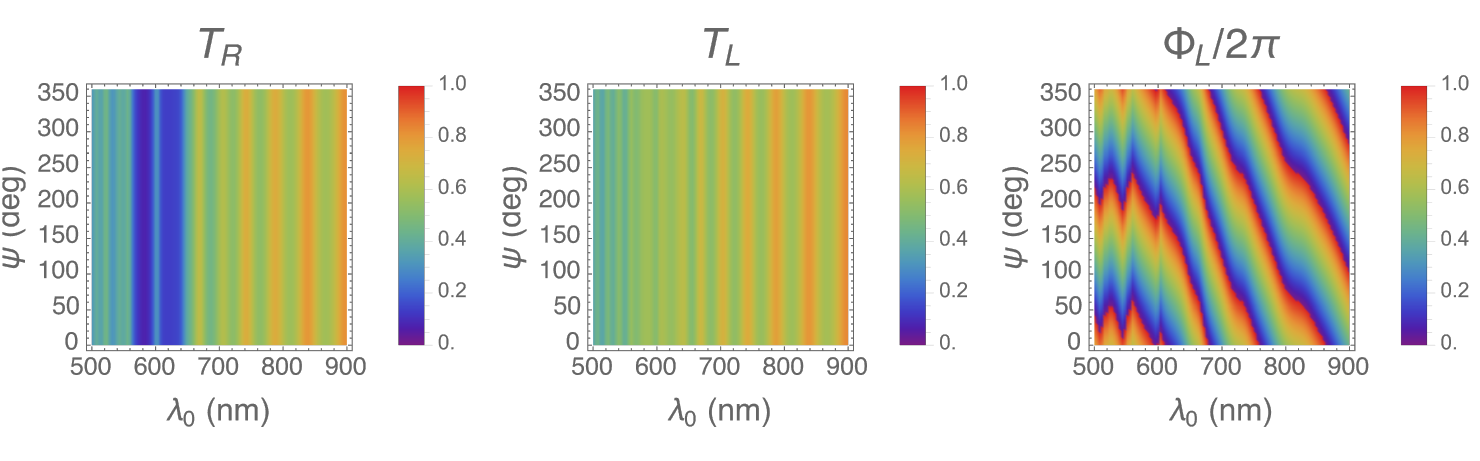}\\
\includegraphics[width=\linewidth]{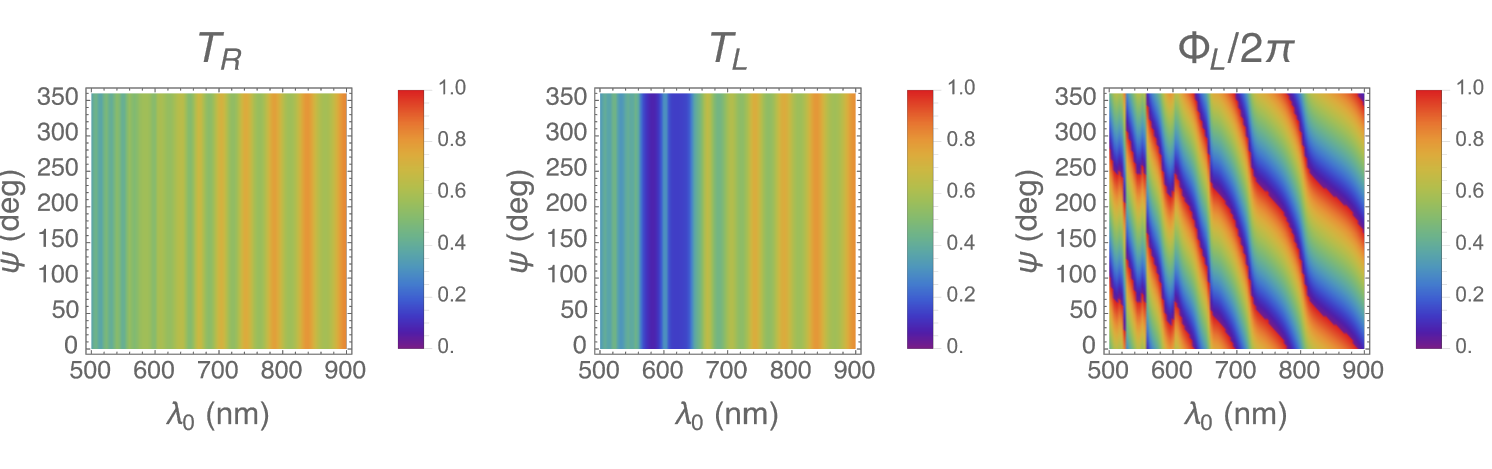}
\end{tabular}
\end{center}
\caption{\label{CombinedDefect-psi}
$\TR$,  $\TL$, and $\bPhiL$ as functions
of $\lambdao$ and $\psi$ of a DSCM slab with  central layer and twist defects ($\LCTF=\Omega/4$ and $\twist=135\deg$),
calculated for $\thetainc=0\deg$,  $\Omega=150$~nm,  and $N=5$. Other parameters are as follows:
$p_{\rm a} = 2.3$, $p_{\rm b} =3.0$, $p_{\rm c} =2.2 $, $\lambda_{\rm a} = \lambda_{\rm c} =260$~nm, $\lambda_{\rm b} = 270$~nm,   and
$N_{\rm a} = N_{\rm b} =N_{\rm c}=130$.
Top row: $h = 1$. Bottom row: $h=-1$.
}
\end{figure}

Figure~\ref{CombinedDefect-psi} presents $\TR$,  $\TL$, and $\bPhiL$ as functions
of $\lambdao\in[400~{\rm nm},900~{\rm nm}]$ and $\psi\in[0\deg,360\deg)$ for a DSCM slab with central layer and twist defects,
when $\thetainc=0\deg$. Clearly, $\TR$ and $\TL$ depend very weakly on $\psi$,
whether in or out of the circular Bragg regime;  and even the spectral reflection hole 
varies very little with $\psi$.
Although the spectral reflection hole is clearly evident as a vertical feature
in the plots of $\bPhiL$, that quantity depends strongly on $\psi$ inside as
well as outside the circular Bragg regime.

Although transmittance data for $\psi\in[180\deg, 360\deg]$ can be obtained from transmittance
data for $\psi\in[180\deg, 360\deg]$ by exploiting symmetry, the same cannot be said for geometric-phase data.
Linearity permits the representation
\begin{equation}
\left.\begin{array}{l}
\tL = \tLL\,\aL+\tLR\,\aR
\\[5pt]
\tR=\tRL\,\aL+\tRR\,\aR
\end{array}\right\}\,,
\end{equation}
so that
$\TL = \vert\tLL\vert^2 +\vert\tRL\vert^2$
and $\TR = \vert\tRR\vert^2 +\vert\tLR\vert^2$.
Examination of calculated data reveals that
\begin{equation}
\left.\begin{array}{l}
\tLL(h,\thetainc,\psi)=\tRR(-h,\thetainc,2\pi-\psi)
\\[5pt]
\tLR(h,\thetainc,\psi)=\tRL(-h,\thetainc,2\pi-\psi)
\\[5pt]
 \tRL(h,\thetainc,\psi)=\tLR(-h,\thetainc,2\pi-\psi)
\\[5pt]
\tRR(h,\thetainc,\psi)=\tLL(-h,\thetainc,2\pi-\psi)
\end{array}\right\}\,.
\label{t-symm}
\end{equation}
These symmetries imply that
\begin{equation}
\left.\begin{array}{l}
\TL(h,\thetainc,\psi)= \TR(-h,\thetainc,2\pi-\psi)
\\[5pt]
\TR(h,\thetainc,\psi)= \TL(-h,\thetainc,2\pi-\psi)
\end{array}\right\}\,,
\label{T-symm}
\end{equation}
so that the transmittances  for $\lec{h,\psi}\ric$ can be obtained from the transmittances
 for $\lec{-h,2\pi-\psi}\ric$. However, the symmetries
\begin{equation}
\left.\begin{array}{l}
\alphatra_{\rm L}(h,\thetainc,\psi)=2\pi-\alphatra_{\rm R}(-h,\thetainc,2\pi-\psi)
\\[5pt]
\alphatra_{\rm R}(h,\thetainc,\psi)=2\pi-\alphatra_{\rm L}(-h,\thetainc,2\pi-\psi)
\end{array}\right\}\,
\end{equation}
and
\begin{equation}
\left.\begin{array}{l}
\betatra_{\rm L}(h,\thetainc,\psi)= -\betatra_{\rm R}(-h,\thetainc,2\pi-\psi)
\\[5pt]
\betatra_{\rm R}(h,\thetainc,\psi)= -\betatra_{\rm L}(-h,\thetainc,2\pi-\psi)
\end{array}\right\}\,,
\end{equation}
which also stem from Eqs.~(\ref{t-symm}),
are unfruitful for a similar exercise with geometric phases because $\bPhiR\equiv0$ but $\bPhiL$
can be non-zero.

By choosing $\LCTF$ and $\twist$ appropriately, the spectral reflection holes can be positioned anywhere
inside the circular regime. For example, when $\thetainc=\psi=0$, these features are located at $\lambdao\approx 610$~nm when
$\LCTF=0.4\Omega$ and $\twist=120\deg$ (results not shown), instead of at $\lambdao\approx 600$~nm when
$\LCTF=0.5\Omega$ and $\twist=135\deg$ (Fig.~\ref{CombinedDefect-psi}).

\subsection{Crossover to spectral transmission holes}

Both types of central phase defects can engender, singly \cite{Kopp,Becchi,WL2005} as well as
jointly \cite{Schmidtke-EPJE,Hodg-PRL}, spectral transmission holes, when $N$
is sufficiently large. A remarkable crossover, from 
\begin{itemize}
\item a spectral reflection hole  in
the response of a structurally right-handed ($h=1$) and non-dissipative
DSCM slab with a central 90-$\deg$
twist defect to a normally incident (i.e., $\thetainc=0$) RCP plane wave \cite{HWTLM-twist}
to
\item a spectral transmission hole in the response
of the same slab  to a normally incident  LCP plane wave
\end{itemize}
with increasing $N$, 
emerged from theoretical analysis \cite{Kopp}. Theory shows  
the analogous crossover, from 
\begin{itemize}
\item a spectral reflection hole  in
the response of a structurally left-handed ($h=-1$) and non-dissipative
DSCM slab with a central 90-$\deg$
twist defect to a normally incident (i.e., $\thetainc=0$) LCP plane wave  
to
\item a spectral transmission hole in the response
of the same slab  to a normally incident  RCP plane wave,
\end{itemize}
with increasing $N$. The spectral reflection hole is considerably wider than the spectral transmission
hole, as explained by coupled-mode theory \cite{WL2005}.

\begin{figure}[!ht]
\begin{center}
\begin{tabular}{c}
\includegraphics[width=0.5\linewidth]{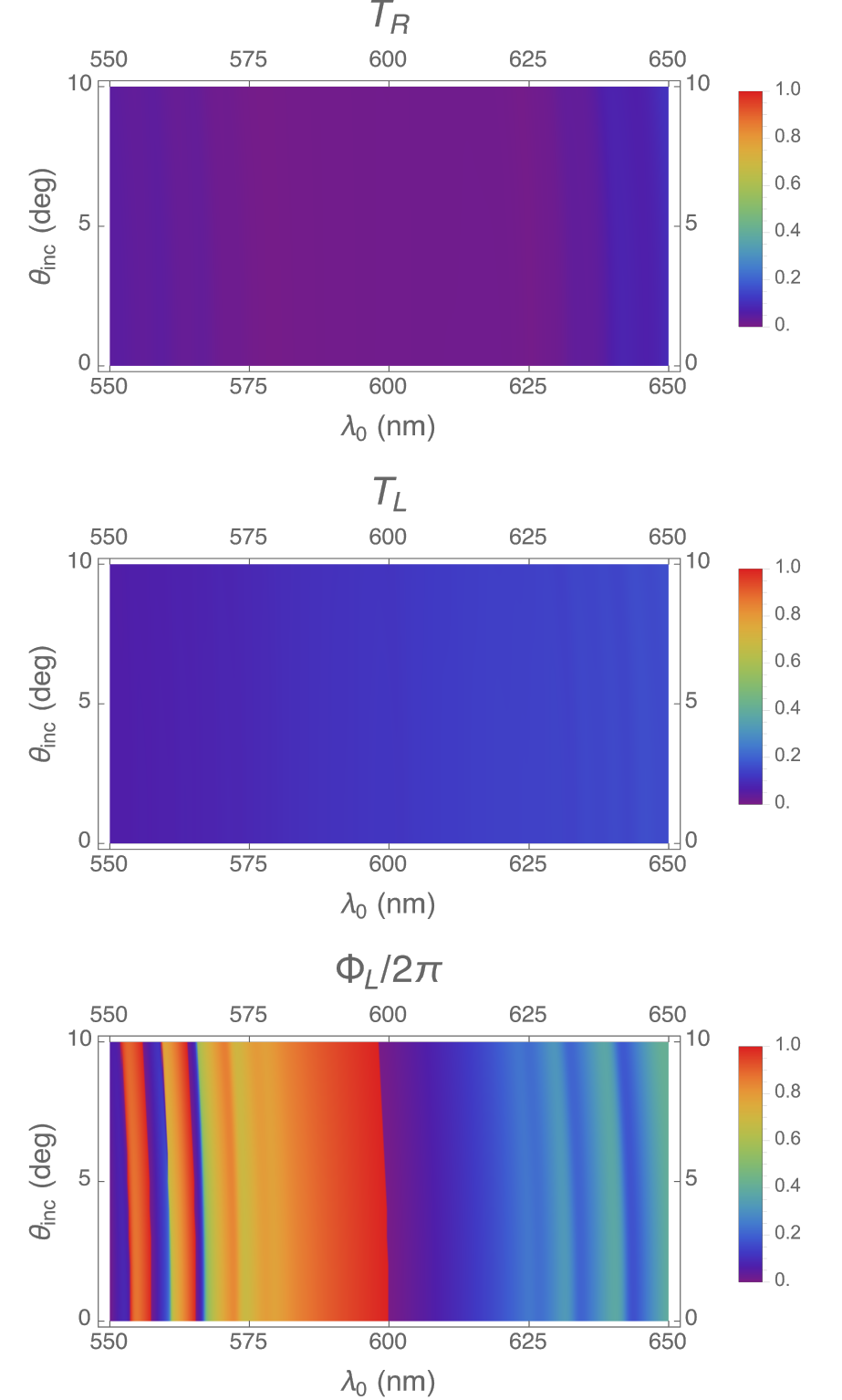}
\includegraphics[width=0.5\linewidth]{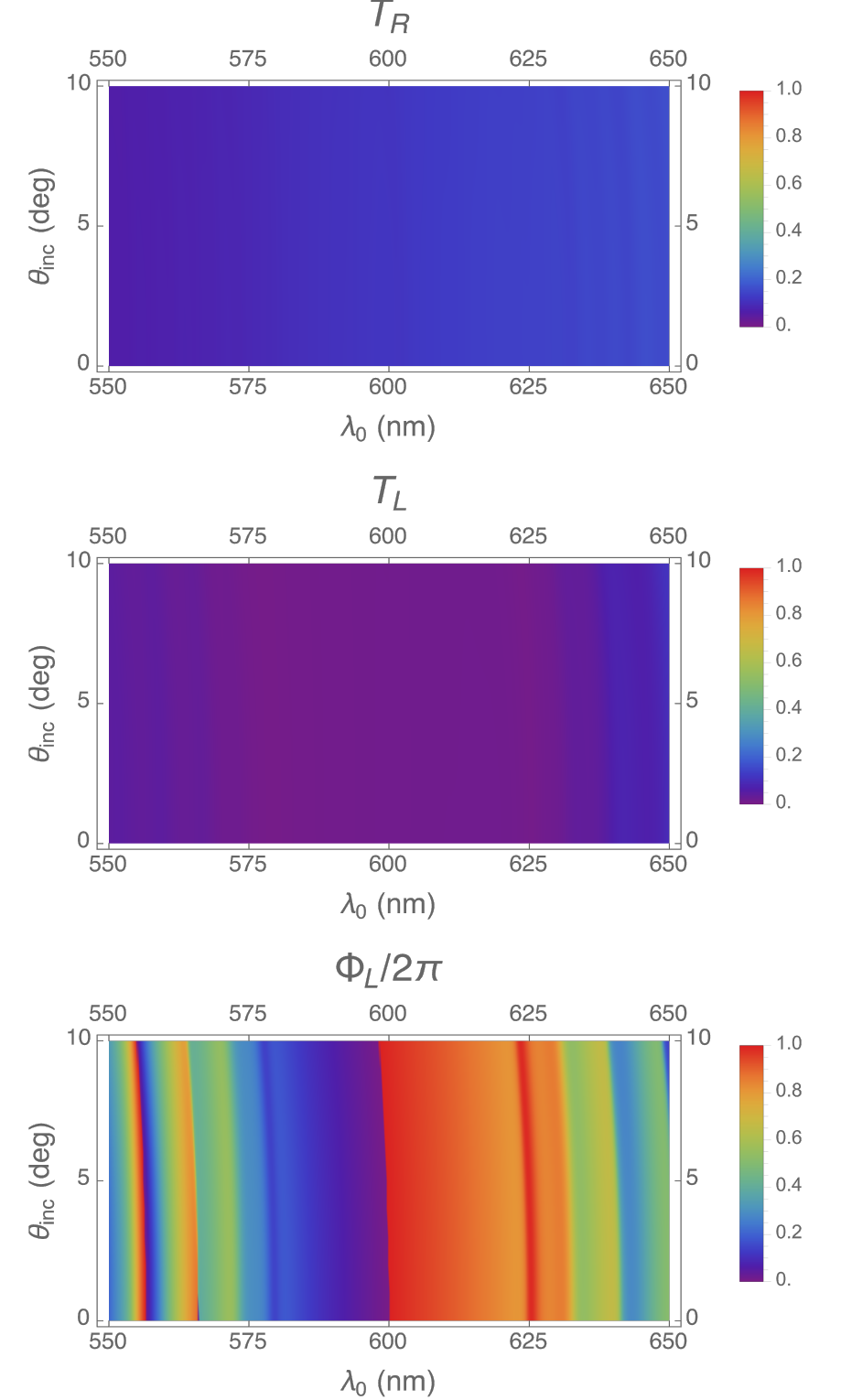}
\end{tabular}
\end{center}
\caption{\label{CombinedDefect2}
$\TR$,  $\TL$, and $\bPhiL$ as functions
of $\lambdao$ and $\thetainc$ of a DSCM slab with  central layer and twist defects ($\LCTF=\Omega/4$ and $\twist=135\deg$),
calculated for $\psi=0\deg$,  $\Omega=150$~nm,  and $N=25$. Other parameters are as follows: $p_{\rm a} = 2.3$, $p_{\rm b} =3.0$, $p_{\rm c} =2.2 $, $\lambda_{\rm a} = \lambda_{\rm c} =260$~nm, $\lambda_{\rm b} = 270$~nm,   and
$N_{\rm a} = N_{\rm b} =N_{\rm c}=130$.
Left column: $h = 1$. Right column: $h=-1$.
}
\end{figure}

Figure~\ref{CombinedDefect2}  is the counterpart of Fig.~\ref{CombinedDefect}, the former drawn for
$N=25$ and the latter for $N=5$. The  reflection hole in the spectrum of $\TR$ (resp. $\TL$)
for $h=1$ (resp. $h=-1$) in  Fig.~\ref{CombinedDefect} has been replaced by a much narrower transmission
hole  in the spectrum of $\TL$ (resp. $\TR$)
for $h=1$ (resp. $h=-1$) in  Fig.~\ref{CombinedDefect2}, whether the plane wave is incident normally or obliquely.

\begin{figure}[!ht]
\begin{center}
\begin{tabular}{c}
\includegraphics[width=0.5\linewidth]{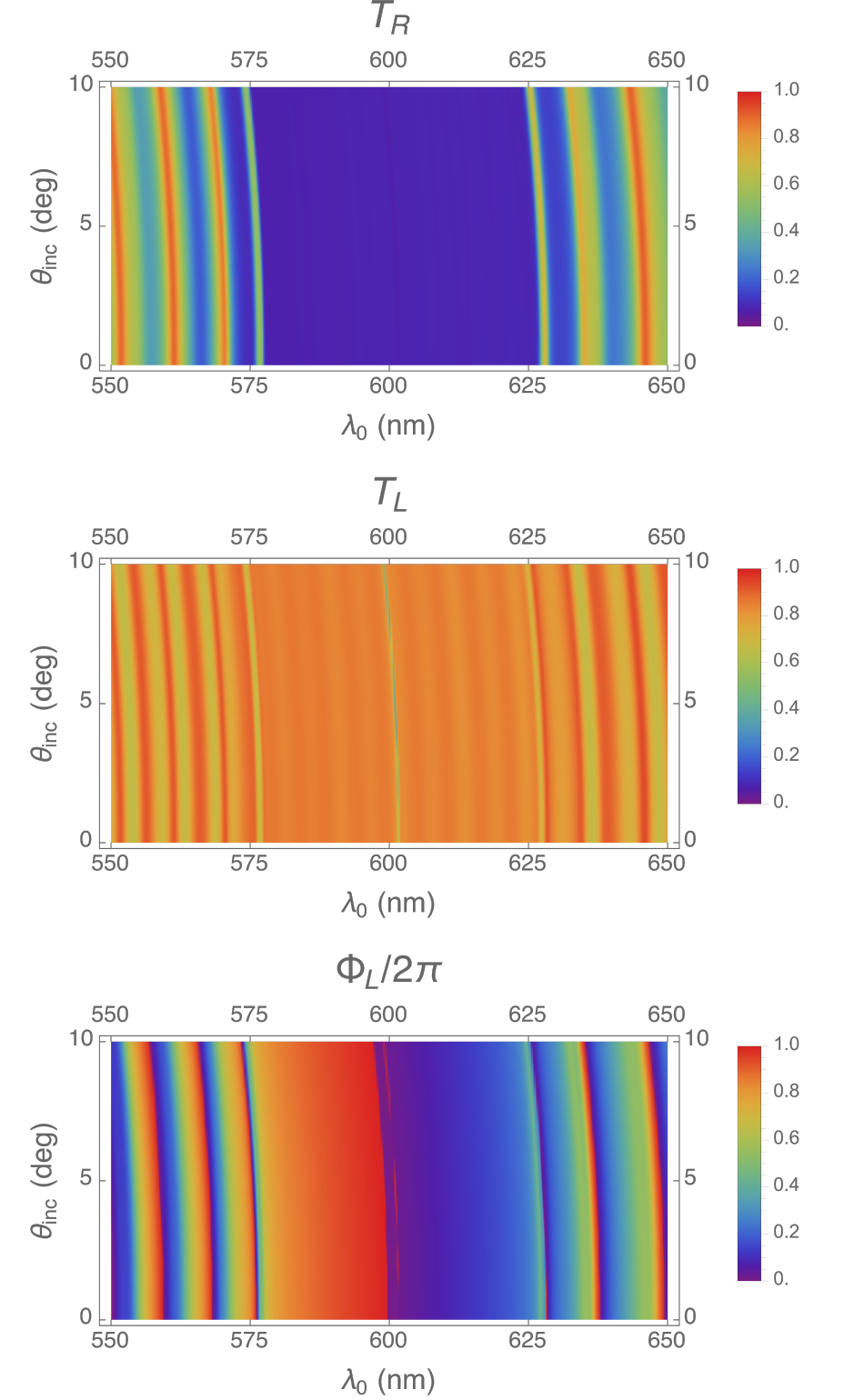}
\includegraphics[width=0.5\linewidth]{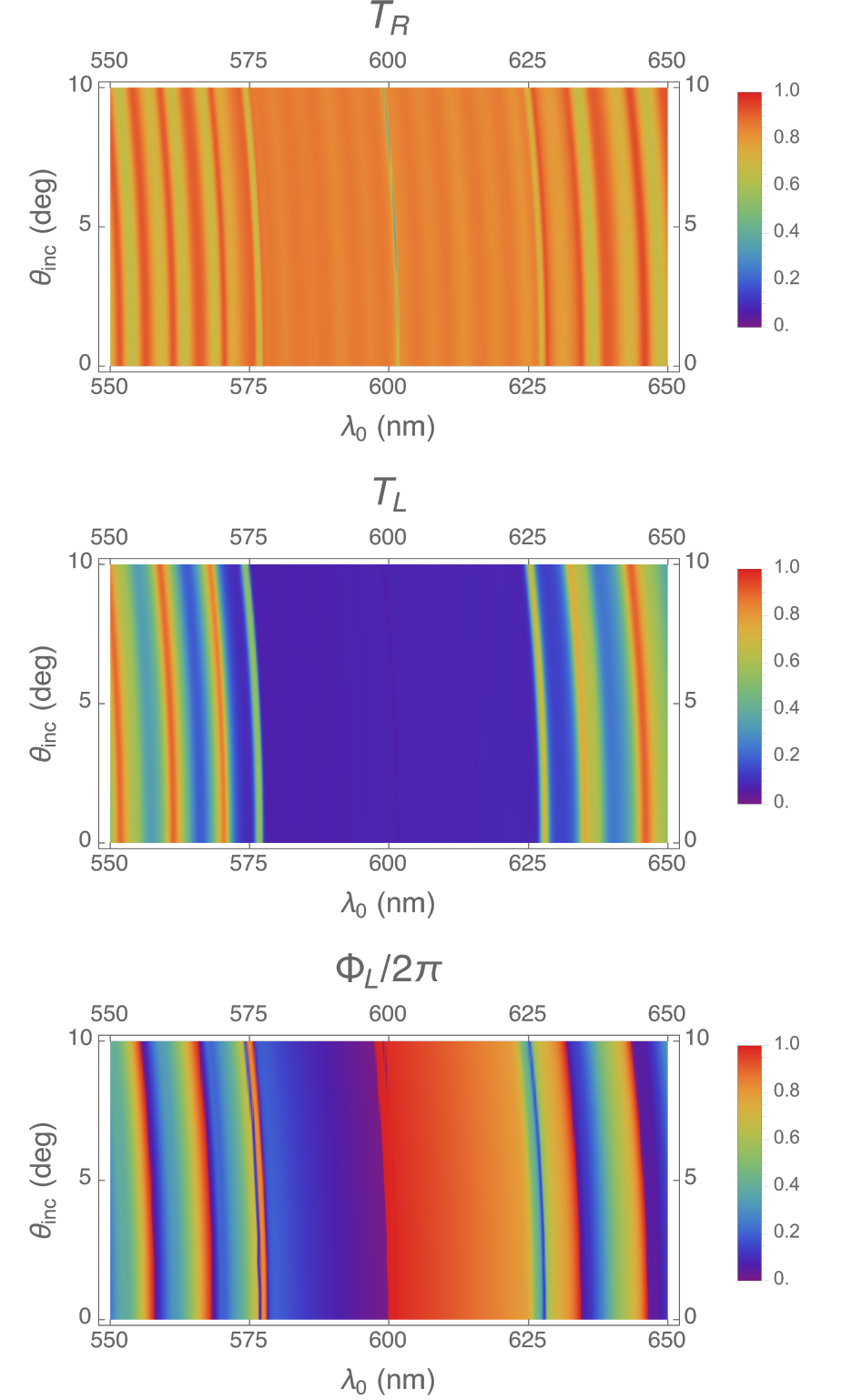}
\end{tabular}
\end{center}
\caption{\label{CombinedDefect3}
$\TR$,  $\TL$, and $\bPhiL$ as functions
of $\lambdao$ and $\thetainc$ of a DSCM slab with  central layer and twist defects ($\LCTF=\Omega/4$ and $\twist=135\deg$),
calculated for $\psi=0\deg$,  $\Omega=162$~nm,  and $N=25$. Other parameters are as follows: $p_{\rm a} = 2.3$, $p_{\rm b} =3.0$, $p_{\rm c} =2.2 $, $\lambda_{\rm a} = \lambda_{\rm b} =\lambda_{\rm c} =10$~nm, and
$N_{\rm a} = N_{\rm b} =N_{\rm c}=130$.
Left column: $h = 1$. Right column: $h=-1$.
}
\end{figure}

Although the ultranarrow spectral transmission hole is clearly present in the plots of $\bPhiL$, it is hard to recognize
it in the transmittance spectrums in Fig.~\ref{CombinedDefect2}, because the much thicker DSCM slab absorbs
electromagnetic radiation very well. When the dissipation in the DSCM was reduced by setting
$\lambda_{\rm a}=\lambda_{\rm b}=\lambda_{\rm c}=10$~nm and calculations were carried out for $N=25$,
the resulting plot of $\TL$ (resp. $\TR$)
for $h=1$ (resp. $h=-1$) in Fig.~\ref{CombinedDefect3} shows the spectral transmission hole very well.

\section{Concluding Remarks}\label{sec:cr}

The effect of inserting a central phase defect in a DSCM slab with a modest number
of structural periods is the emergence
of a narrowband high-transmittance feature (i.e., a spectral reflection hole)
in the circular Bragg regime, only when the handedness of the incident circularly polarized plane
wave is the same as the structural handedness of the DSCM. 
However, regardless
of the structural handedness of the DSCM, the geometric phase
of the  transmitted plane wave contains evidence of both the circular Bragg regime and the
spectral reflection hole, if the incident plane wave is LCP. The geometric phase of the transmitted
plane wave is identically zero,  if the incident plane wave is RCP.

A comparison of Figs.~\ref{TwistDefect}--\ref{CombinedDefect} indicates that
the spectral reflection holes due to a central twist defect, a central layer defect,
or a combined defect in a DSCM slab manifest in the same way in the transmittance plots. However,
the geometric-phase signatures of both types of defects and of their combination
are all different. Thus, the type of central phase defect could conceivably
be gleaned by determining $\bPhiL$ as a function of
$\lambdao$ and $\thetainc$, possibly using machine-learning techniques \cite{Bishop,Fieguth}.

When the DSCM slab with the central phase defect is thick enough to have a large number of the structural periods, the
narrowband high-transmittance feature in the response to an incident co-handed circularly polarized plane wave
is replaced by an ultranarrowband high-reflectance feature in the response to an incident cross-handed circularly polarized plane wave.
The new feature may be difficult to observe experimentally because of absorption inside the DSCM slab, but it will still be
evident in  the geometric phase
of the  transmitted plane wave, if the incident plane wave is LCP.

 \vspace{5mm}
\noindent {\bf Acknowledgment.} 
The author  thanks Ricardo A. Fiallo for assistance with a figure
and the Charles Godfrey Binder Endowment at Penn State
  for continued support of his research.

\end{document}